\documentclass[aps,prb,showpacs,preprint,groupedaddress]{revtex4}
\usepackage{graphicx}

\begin{document}


\title{Resistivity, Hall effect and Shubnikov-de Haas oscillations in CeNiSn}


\author{T. Terashima}
\author{C. Terakura}
\author{S. Uji}
\affiliation{National Institute for Materials Science, Tsukuba, Ibaraki 305-0003, Japan}

\author{H. Aoki}
\affiliation{Center for Low Temperature Science, Tohoku University, Sendai, Miyagi 980-8578, Japan}

\author{Y. Echizen}
\author{T. Takabatake}
\affiliation{Department of Quantum Matter, ADSM, Hiroshima University, Higashi-Hiroshima 739-8530, Japan}

\date{\today}

\begin{abstract}
The resistivity and Hall effect in CeNiSn are measured at temperatures down to 35 mK and in magnetic fields up to 20 T with the current applied along the {\it b} axis.  The resistivity at zero field exhibits quadratic temperature dependence below $\sim$0.16 K with a huge coefficient of the $T^2$ term (54 $\mu$$\Omega$cm/K$^2$).  The resistivity as a function of field shows an anomalous maximum and dip, the positions of which vary with field directions.  Shubnikov-de Haas (SdH) oscillations with a frequency {\it F} of $\sim$100 T are observed for a wide range of field directions in the {\it ac} and {\it bc} planes, and the quasiparticle mass is determined to be $\sim$10-20 {\it m}$_e$.  The carrier density is estimated to be $\sim$10$^{-3}$ electron/Ce.  In a narrow range of field directions in the {\it ac} plane, where the magnetoresistance-dip anomaly manifests itself clearer than in other field directions, a higher-frequency ($F=300\sim400\text{ T}$) SdH oscillation is found at high fields above the anomaly.  This observation is discussed in terms of possible field-induced changes in the electronic structure.
\end{abstract}


\maketitle

\section{Introduction}
Kondo semiconductors (or insulators), e.g., SmB$_6$, YbB$_{12}$, and Ce$_3$Bi$_4$Pt$_3$, are cousins of heavy-fermion compounds.\cite{Takabatake98, Aeppli92}  Both classes of materials behave like Kondo-impurity systems at high temperatures, {\it f}-electron spins independently scattering conduction electrons.  As the scattering becomes coherent at low temperatures, the heavy-fermion compounds remain metallic, described as a Fermi liquid of heavy quasiparticles, while the Kondo semiconductors become semiconducting with a small energy gap opening in the excitation spectrum.  Since the known Kondo semiconductors have an even number of electrons per unit cell, it has been suggested that, despite strong electron-correlation effects, they may be viewed as band insulators.\cite{Aeppli92}  Namely, if one occupied {\it f} state hybridizes with a single half-filled conduction band, a band insulator may result, with the lower hybridized band completely occupied.

CeNiSn, crystallizing in an orthorhombic structure, is a unique variety of the Kondo semiconductors, in which the gap formation seems incomplete.\cite{Takabatake98}  The Kondo temperature {\it T}$_K$ of 51 K is deduced from specific heat data above $\sim$10 K,\cite{Nishigori96} which is comparable to {\it T}$_K$ of 24 K in the representative heavy-fermion compound CeRu$_2$Si$_2$.\cite{Besnus85}  Although the semiconducting behavior of resistivity observed in early samples (Ref.~\onlinecite{Takabatake90}) is completely suppressed as the quality of crystals is improved,\cite{Nakamoto95} there is plenty of evidence for the gap formation below $\sim$10 K, e.g., direct observations of the gap by break-junction tunneling spectroscopy (Ref.~\onlinecite{Ekino95}) and the rapid suppression of {\it C}/{\it T} (Refs.~\onlinecite{Nishigori96, Izawa96}) and 1/{\it T}$_1$,\cite{Kyogaku90} where {\it C} is the electronic part of specific heat, {\it T} temperature and {\it T}$_1$ the nuclear spin lattice relaxation time.  However, {\it C}/{\it T} levels off at $\sim$40 mJ/molK$^2$ below 1 K,\cite{Izawa99} and 1/{\it T}$_1$ exhibits the Korringa law ({\it T}$_1${\it T} = constant) below 0.4 K.\cite{Nakamura94}  Further, the thermal conductivity exhibits {\it T}-linear dependence below 0.3 K,\cite{Paschen00} which is characteristic of metallic systems as {\it T} $\rightarrow$ 0.  Analyses of {\it C}/{\it T} and 1/{\it T}$_1$ suggest that the gap is a V-shaped pseudogap of the width $\sim$10 K with finite density of states (DOS) at the Fermi level.\cite{Izawa96, Kyogaku90, Nakamura94}  Since the width of the gap is small, the magnetic field {\it B} may alter the electronic structure via the Zeeman splitting of the up- and down-spin energy bands.  There are some experimental indications that the gap may be suppressed by the field applied along the {\it a} axis,\cite{Takabatake92, Inada96, Izawa96, Sugiyama98, Terashima01} which is the easy axis of magnetization.

On the theoretical side, there are two contrasting approaches to CeNiSn.  One is an anisotropic hybridization-gap model.\cite{Ikeda96, Moreno00}  It is argued that, if the lowest Kramers doublet of the Ce 4{\it f} state has special symmetry, the hybridization between conduction and {\it f} electrons and hence the gap may vanish in some directions in the Brillouin zone, leading to finite DOS at the Fermi level.  The other approach takes a view of a two-fluid model, assuming the existence of neutral spin excitations that are decoupled from charged Fermi-liquid excitations (i.e., charge carriers).\cite{Kagan97}  It is claimed that the spin excitations dominate low-temperature thermodynamic properties and that a pseudogap opens only in their spectrum.

In a previous paper, we have reported the first observation of Shubnikov-de Haas (SdH) oscillations in CeNiSn.\cite{TerashimaSCES}  In this paper, we present more extensive measurements of low-temperature resistivity and Hall effect as well as SdH oscillations in CeNiSn.  We discuss the electronic structure and the influence of the magnetic field on it.

\section{Experiments}
Single-crystalline ingots of CeNiSn were grown by the Czochralski method and were purified by a solid-state electrotransport technique.\cite{Nakamoto95}  Two parallelpipeds (samples-b1 and -b2) of $\sim$0.5 mm$^2$ in cross section and $\sim$4 mm in length along the {\it b} axis were cut from an ingot by spark erosion.  Gold lead wires were spot-welded.  The resistivity and Hall effect were measured at temperatures down to 35 mK and in magnetic fields up to 20 T by a conventional four-terminal method with a low-frequency ac excitation current ({\it f} = 17 Hz, {\it I} = 100 $\mu$A) applied along the {\it b} axis.  The field was rotated in the {\it ac} and {\it bc} planes, and the field angles $\theta$$_{ca}$ and $\theta$$_{cb}$ are measured from the {\it c} axis.  Both samples were metallic down to the lowest temperature.

\section{Results}

\begin{figure}
\includegraphics[width=8.5cm]{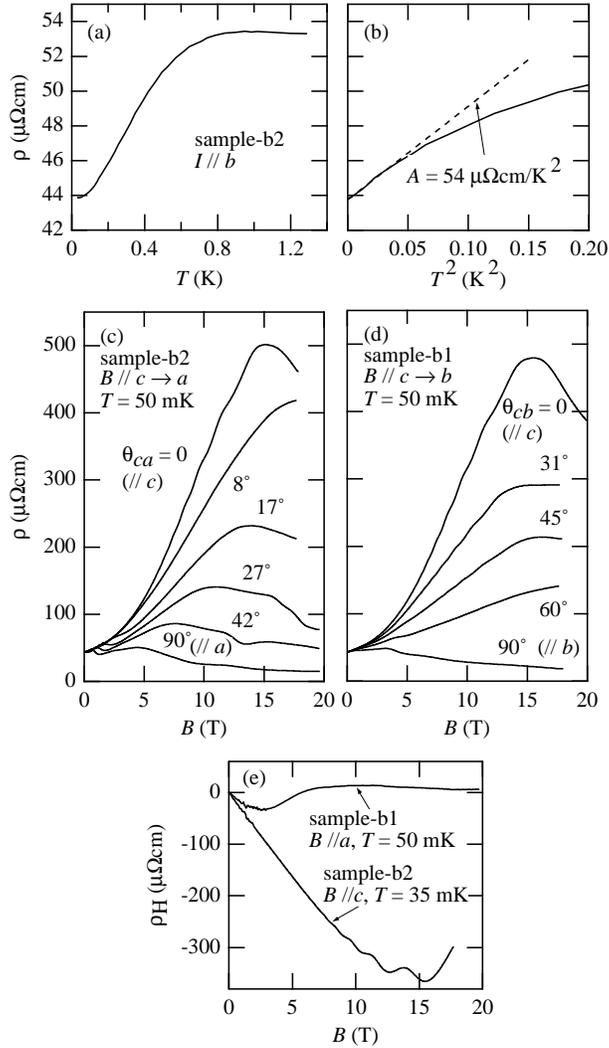}
\caption{\label{fig:1}(a) Resistivity $\rho$ in CeNiSn measured along the {\it b} axis as a function of temperature {\it T}.  (b) Low temperature part of the same data as a function of {\it T}$^2$.  The broken line shows a fit to $\rho$ =$\rho$$_o$ + {\it A}{\it T}$^2$ below 0.16 K.  (c) and (d) $\rho$ vs field {\it B} for selected field directions in the {\it ac} and {\it bc} planes, respectively.  The field angles $\theta$$_{ca}$ and $\theta$$_{cb}$ are measured from the {\it c} axis.  Note that the field geometry is always transverse in (c), while it changes from transverse at $\theta$$_{cb}$ = 0 to longitudinal at $\theta$$_{cb}$ = 90$^{\circ}$ in (d).  (e) Hall resistivity $\rho$$_H$ vs {\it B} for {\it B} $\parallel$ {\it a} and {\it B} $\parallel$ {\it c}.}
\end{figure}

Figure~\ref{fig:1}(a) shows the temperature dependence of resistivity, which is quite unusual.  The resistivity is nearly constant around 1 K, as if a residual resistivity regime is reached, yet it starts to decrease again near 0.8 K.  The resistivity exhibits a quadratic temperature dependence at the lowest temperatures [Fig.~\ref{fig:1}(b)], which is a characteristic of a Fermi liquid.  The coherence temperature $T_{\text{coh}}$, i.e., the upper limit of the {\it T}$^2$ behavior, is very low, $\sim$0.16 K.  The coefficient {\it A} of the {\it T}$^2$ term is huge, 54 $\mu$$\Omega$cm/K$^2$.  These points may be illustrated by comparison with the corresponding values found in CeRu$_2$Si$_2$, i.e., $T_{\text{coh}}$ $\sim$ 0.6 K and {\it A} = 0.4 $\mu$$\Omega$cm/K$^2$.\cite{Mignot89} 

Figure~\ref{fig:1}(c) shows the magnetoresistance for selected field directions in the {\it ac} plane.  The magnetoresistance for {\it B} $\parallel$ {\it a} ($\theta$$_{ca}$ = 90$^{\circ}$) exhibits a broad maximum at 4.4 T and then decreases, showing a bend near 9 T.  This behavior is in accord with previous data at 0.45 K.\cite{Inada96}  Note that the field geometry is transverse.  Since CeNiSn has an even number of electrons per unit cell and hence must be a compensated metal, the negative magnetoresistance at high fields is unexpected.  The position of the magnetoresistance maximum moves to higher fields with decreasing $\theta$$_{ca}$ (e.g., 7.6 T at $\theta$$_{ca}$ = 42$^{\circ}$), going beyond the investigated field range near $\theta$$_{ca}$ = 8$^{\circ}$, and then comes back to 15 T for {\it B} $\parallel$ {\it c} ($\theta$$_{ca}$ = 0).  The bend also shifts to higher fields with decreasing $\theta_{ca}$ and evolves into a dip for $\theta$$_{ca}$ $<$ $\sim$60$^{\circ}$, as seen at 13.5 T at $\theta$$_{ca}$ = 42$^{\circ}$.  As the field is further rotated, the dip is preceded by an extended region where the resistivity exhibits rapid decrease; at $\theta$$_{ca}$ = 27$^{\circ}$, the rapid decrease starts at 15.5 T, but the dip is not reached up to 20 T.  For {\it B} $\parallel$ {\it c}, the oscillatory behavior below 15 T is due to the SdH effect and will be described in more detail below.

Figure~\ref{fig:1}(d) shows the magnetoresistance for selected field directions in the {\it bc} plane.  The field geometry changes from transverse at $\theta$$_{cb}$ = 0 to longitudinal at $\theta$$_{cb}$ = 90$^{\circ}$.  Although this leads to much suppressed magnetoresistance for large $\theta$$_{cb}$, the position of the magnetoresistance maximum can be traced.  As the field is tilted from the {\it c} axis with increasing $\theta$$_{cb}$, the maximum position first moves to slightly lower fields, then goes up, and appears to go out of the field range near $\theta$$_{cb}$ = 60$^{\circ}$.

The Hall resistivity for {\it B} $\parallel$ {\it a} is also anomalous [Fig.~\ref{fig:1}(e)]; it exhibits a negative peak at 2.5 T and takes small positive values at high fields.  Similar behavior was previously observed at 1.5 K.\cite{Takabatake96}  For {\it B} $\parallel$ {\it c}, the Hall resistivity is linear in {\it B} at low fields and then bends near 15 T in line with the anomaly in the magnetoresistance.  The oscillatory behavior below 15 T is ascribed to the SdH effect.  Although CeNiSn must have both electron and hole carriers because of the charge compensation, the large negative Hall resistivity suggests that electron carriers dominate the electrical conduction.  Accordingly, if we assume a single-carrier model, the Hall coefficient at 2 T, -33$\times$10$^{-2}$ cm$^3$/C, corresponds to the carrier concentration of 1.2$\times$10$^{-3}$ electron/Ce.

\begin{figure}
\includegraphics[width=8.5cm]{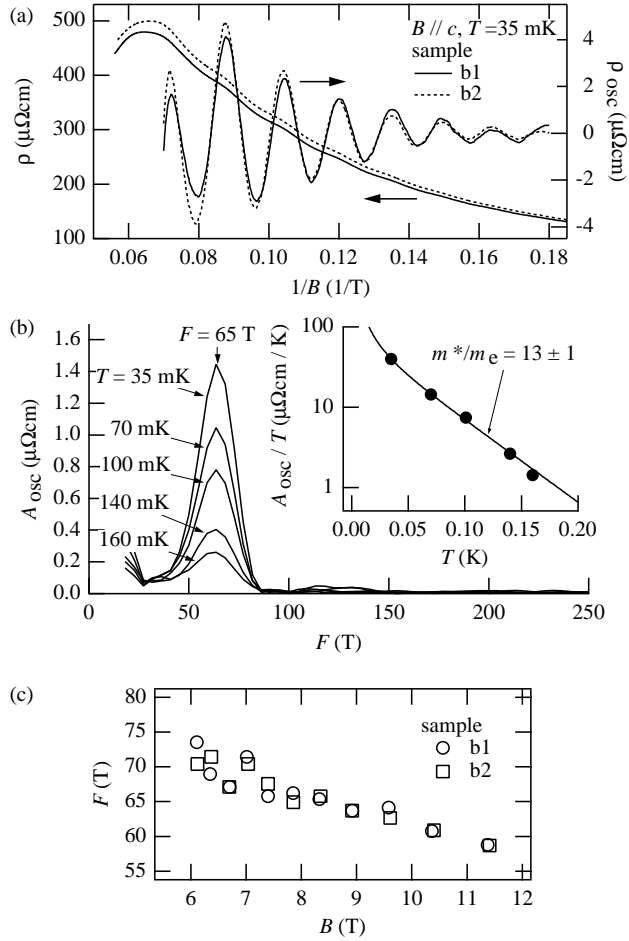}
\caption{\label{fig:2}(a) Resistivity $\rho$ in CeNiSn for the field parallel to the {\it c} axis and its oscillatory part $\rho_{\text{osc}}$ as functions of the inverse field 1/{\it B}.  To obtain $\rho_{\text{osc}}$, a polynomial was fitted to the smoothly varying background of $\rho$(1/{\it B}) and subtracted from it.  The two samples, b1 and b2, give essentially identical results.  (b) The Fourier transform of $\rho_{\text{osc}}$ in the sample b1.  Spectra at higher temperatures are also shown.  The inset shows the temperature dependence of SdH oscillation amplitudes $A_{\text{osc}}$.  The solid curve is a fit to the Lifshitz-Kosevich formula,\cite{Shoenberg84} which yields the effective mass $m^*$ of 13$\pm$1 in the units of the free electron mass {\it m}$_e$.  (c) The frequency determined from each one oscillation period (peak-to-peak or valley-to-valley width) as a function of the field for the two samples.}
\end{figure}

In Fig.~\ref{fig:2}(a), we show the oscillatory part of the magnetoresistance measured with {\it B} $\parallel$ {\it c} for the two samples.  We see that both samples exhibit essentially identical SdH oscillations.  The Fourier transform of the oscillations indicates a single frequency at 65$\pm$5 T [Fig.~\ref{fig:2}(b)], where the error is simply set by the inverse of the 1/{\it B}-width of the original data.  The corresponding orbit area is 0.5$\%$ of the cross-section normal to the {\it c} axis of the Brillouin zone.  Fitting the temperature dependence of the oscillation amplitude to the Lifshitz-Kosevich formula \cite{Shoenberg84} (inset), we find the associated effective mass to be 13$\pm$1 {\it m}$_e$, where {\it m}$_e$ is the free electron mass.

Although the Fourier transforms show a single peak, a closer examination of the oscillations reveals that the frequency actually changes with field.   Figure~\ref{fig:2}(c) shows the frequency determined from each one oscillation period (peak to peak or valley to valley) as a function of field.  We find that the frequency changes from $\sim$70 to $\sim$60 T with increasing field.  We also examined possible field dependence of the effective mass for the sample b1 by measuring a peak-to-valley height of each oscillation.  The masses thus determined are 12.3, 13.2, 14.2, and 13.9 $m_e$ at 7.4, 8.4, 9.6, and 11.5 T, respectively.  It appears that the mass slightly increases with field.  However, since the errors associated with these estimations are $\pm$1 $m_e$, more precise measurements are necessary to conclusively state the field dependence of the mass.

\begin{figure}
\includegraphics[width=8.5cm]{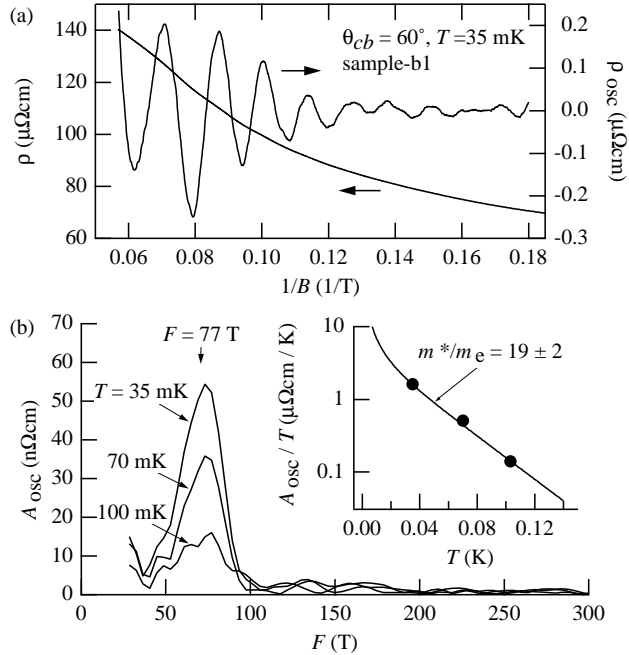}
\caption{\label{fig:3}(a) Resistivity $\rho$ in CeNiSn for the field angle $\theta$$_{cb}$ = 60$^{\circ}$ and its oscillatory part $\rho_{\text{osc}}$ as functions of the inverse field 1/{\it B}.  (b) The Fourier transform of $\rho_{\text{osc}}$.  Spectra at higher temperatures are also shown.  The inset shows the temperature dependence of SdH oscillation amplitudes $A_{\text{osc}}$.  The effective mass $m^*$ is determined to be 19$\pm$2 {\it m}$_e$}
\end{figure}

\begin{figure}
\includegraphics[width=6cm]{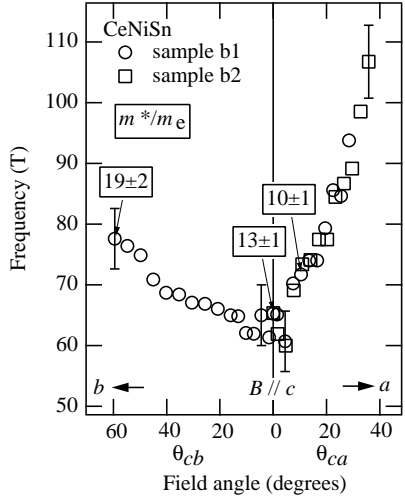}
\caption{\label{fig:4}The field-direction dependence of the SdH frequency in CeNiSn.  The frequencies were determined from Fourier transforms of oscillations in the following field intervals: {\it B} = 5.56 - 14.3 T for all $\theta$$_{cb}$'s and for $\theta$$_{ca}$ $<$ 26$^{\circ}$, and {\it B} = 5.56 - 10.9 T for  $\theta$$_{ca}$ $\geq$ 26$^{\circ}$.  For the {\it ac} plane, the measurements on the two samples yielded identical results within experimental error (compare the circles and squares).  The effective mass is also shown for selected field directions.}
\end{figure}

SdH oscillations of this frequency branch are observed in a wide range of field directions; $\theta$$_{cb}$ up to 60$^{\circ}$ and $\theta$$_{ca}$ up to 36$^{\circ}$.  For instance, we show the oscillatory part of the magnetoresistance measured at $\theta$$_{cb}$ = 60$^{\circ}$ as a function of 1/{\it B} in Fig.~\ref{fig:3}(a).  The Fourier spectrum indicates a frequency of 77$\pm$5 T [Fig.~\ref{fig:3}(b)], and the effective mass is determined to be 19$\pm$2 {\it m}$_e$ (inset).  We plot the angular dependence of the SdH frequency in Fig.~\ref{fig:4}.  Here the frequencies were determined from Fourier transforms of the data in the same field interval from 5.56 to 14.3 T for all $\theta$$_{cb}$'s and for $\theta$$_{ca}$ $<$ 26$^{\circ}$.  For $\theta$$_{ca}$ $\geq$ 26$^{\circ}$, a narrower window from 5.56 to 10.9 T was used so that the magnetoresistance maximum and bend (or dip) will not affect the Fourier transforms.  The frequency increases faster than 1/cos$\theta_{ca}$ in the {\it ac} plane, while the angular dependence in the {\it bc} plane is weak.  As a rough approximation, the shape of the Fermi surface (FS) seems a concave lens of which the axis is parallel to the {\it a} axis.

We here note evidence that the observed SdH oscillations are intrinsic to high-quality single crystals of pure CeNiSn, not due to impurities.  (1) We performed measurements not only at positive field angles ( $\theta$$_{ca}$ or $\theta$$_{cb}$ $>$ 0) but also at negative ones ( $\theta$$_{ca}$ or $\theta$$_{cb}$ $<$ 0).  It was verified that the angular dependences of the SdH frequency and amplitude were symmetric with respect to the {\it c} axis.  Namely, the SdH oscillations exhibit the appropriate symmetry of the crystal.  (2) The angular dependences in the {\it ac} plane determined for the two samples are identical within error (Fig.~\ref{fig:4}).

\begin{figure}
\includegraphics[width=8.5cm]{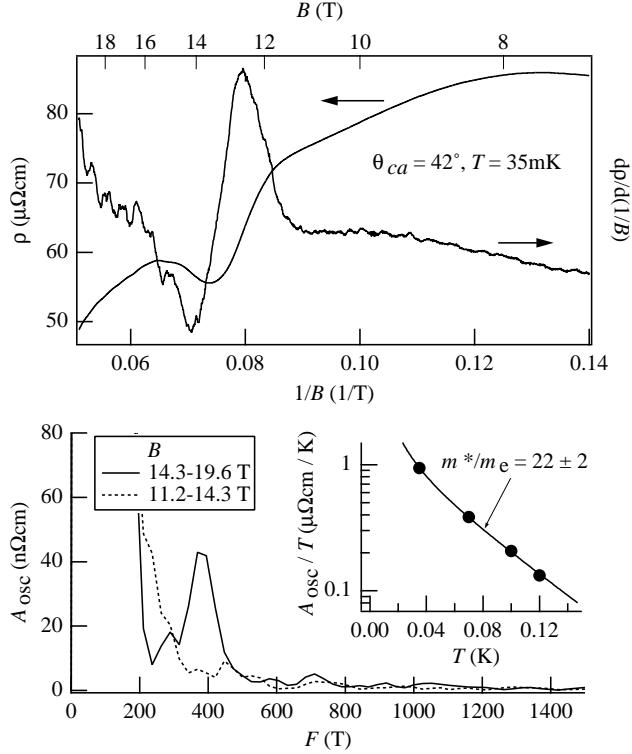}
\caption{\label{fig:5}(a) Resistivity $\rho$ in CeNiSn for the field angle $\theta$$_{ca}$ = 42$^{\circ}$ and its derivative with respect to the inverse field 1/{\it B} as functions of 1/{\it B}.  The magnetoresistance decreases fairly sharply from 12 T and exhibits a dip at 13.5 T.  The derivative curve indicates the existence of SdH oscillations above this dip.  (b) The Fourier spectra for field ranges above (solid curve) and below (broken curve) the magnetoresistance dip.  The inset shows the temperature dependence of the amplitude $A_{\text{osc}}$ of SdH oscillations observed above the dip.}
\end{figure}

For a limited range of field directions, we have observed another frequency branch at high fields above the aforementioned magnetoresistance dip.  Figure~\ref{fig:5}(a) shows the magnetoresistance and its derivative with respect to 1/{\it B} for the field angle $\theta$$_{ca}$ = 42$^{\circ}$ as a function of 1/{\it B}.  The magnetoresistance decreases fairly sharply from $\sim$12 T and exhibits a minimum at 13.5 T.  In the field range higher than this dip, the magnetoresistance exhibits oscillatory behavior, which is more clearly visible in the derivative curve.  The Fourier transform of the magnetoresistance in the field range above the dip [Fig.~\ref{fig:5}(b), solid curve] shows a clear peak at $\sim$380 T, while the spectrum for the field range below the dip (broken curve) does not show any peak.  The effective mass for the orbit is determined to be 22$\pm$2 {\it m}$_e$ (inset).  For the nearby field directions $\theta$$_{ca}$ = 48$^{\circ}$ and 55$^{\circ}$, SdH oscillations with a frequency of $\sim$300 T appear similarly on the high-field side of the magnetoresistance dip.  For still larger field angles, $\theta$$_{ca}$ $>$ 55$^{\circ}$, no oscillations are observed.  This may be attributed to that the resistivity at high fields and hence its oscillatory part are so small for field directions close to the {\it a} axis [see Fig.~\ref{fig:1}(c)].

\section{Discussion}

We begin with the field-dependence of the SdH frequency shown in Fig.~\ref{fig:2}(c).  When a SdH frequency depends on the field, the momentary frequency $F_m$ that is experimentally measured differs from the true frequency $F_{\text{true}}$ that is directly connected with an extremal cross-section of the FS; $F_m=F_{\text{true}}-B(\text{d}F_{\text{true}}/\text{d}B)$.\cite{Ruitenbeek82}  This means that (1) a linear variation in $F_{\text{true}}$ does not affect $F_m$, and that, (2) when $F_{\text{true}}$ has non-linear field dependence, the variation in $F_m$ may be enhanced over that in $F_{\text{true}}$.  A linear Zeeman shift of a parabolic energy band leads only to a linear change in a FS cross-section and hence does not affect $F_m$, i.e., $F_m$'s of oscillations from up- and down-spin electrons remain the same and constant.  The observed field-dependence indicates that the energy band shifts non-linearly with fields and/or that the energy dispersion in the vicinity of the Fermi level strongly deviates from quadratic.  When theoretical models become available for comparison, further information can be extracted from the present observation.

We next discuss the FS responsible for the frequency branch shown in Fig.~\ref{fig:4}.  The size of the observed SdH frequencies, the order of $\sim$100 T, is compatible with the carrier concentration estimated from the Hall coefficient with a single-carrier model; a spherical FS enclosing 10$^{-3}$ electron/Ce would give a SdH frequency of 190 T.  The compatibility supports that what we have found from the SdH oscillations is not small structures of a large FS but the major portion of a small but dominant FS.  Since the Hall resistivity is negative, the observed FS is an electron pocket.   Note that, even if the measured Hall resistivity contains some contribution from hole carriers, the above line of argument is still valid.  In that case, the true carrier number is smaller than the single-carrier-model estimation, and hence it becomes even more unlikely that larger FS's are hidden.

It is interesting to estimate the contribution to $\gamma$ (= {\it C}/{\it T}) of this electron FS.  Using isotropic three-dimensional effective-mass approximation, $\gamma$ is given by $\pi^2k_{\text B}^2n/2\epsilon_F$, where $\epsilon_F=\hbar^2k_F^2/2m^*$ and $k_F^3=3\pi^2n/v$.  Here, $\epsilon_F$ is the Fermi energy, $k_F$ the Fermi wave number, $v$ unit volume, and other symbols as usual.  Assuming {\it n} to be 10$^{-3}$ electron/Ce and $m^*$ to be 10 - 20 $m_e$, we have $\epsilon_F$ of 26 - 13 K and $\gamma$ of 1.5 - 3 mJ/molK$^2$.  The estimated $\gamma$ is one order-of-magnitude smaller than the experimental value of 40 mJ/molK$^2$.\cite{Izawa99}

Having discussed the experimental estimations of the carrier number and $\gamma$, we here compare the two theoretical models of CeNiSn.  The considerable discrepancy between the estimated and measured $\gamma$ values may be a support to the two-fluid model,\cite{Kagan97} since it claims that the specific heat is dominated by neutral spin excitations.  However, it does not seem to provide clear explanation for the largely reduced carrier number.  On the other hand, in the case of the anisotropic hybridization-gap model, small FS's are expected to appear around the nodes in the Brillouin zone where the hybridization gap vanishes, if small dispersion of renormalized {\it f}-electron levels is taken into account.\cite{Ikeda96}  Thus, the observation of the FS does not conflict with the model.  The measured large $\gamma$ may be accounted for if we assume that hole carriers have effective mass of $\sim$200 {\it m}$_e$.  This assumption is not so unlikely as it may, at first sight, appear.  It is known that the many-body mass enhancement in Ce-based heavy fermions varies from FS to FS and hence that light and heavy carriers often coexist in one compound.  Indeed, effective masses in CeRu$_2$Si$_2$ found by de Haas-van Alphen measurements range from $\sim$1 {\it m}$_e$ to more than 100 {\it m}$_e$.\cite{Aoki93}  If {\it T}$_K$ is a measure of possible mass enhancement factor, the existence of carriers with mass of $\sim$200 {\it m}$_e$ in CeNiSn can not be ruled out, since {\it T}$_K$ is comparable in the two compounds.

We now turn to the $T^2$ dependence of resistivity observed below 0.16 K [Fig.~\ref{fig:1}(b)].  We attribute it to quasiparticle-quasiparticle collisions.  Namely, we regard it as a low-temperature limiting behavior of a Fermi liquid.  The two peculiarities in the {\it T}$^2$ dependence, i.e., the relatively low $T_{\text{coh}}$ and anomalously large {\it A} can not be taken as indications of exceptionally strong many-body effects in CeNiSn, but rather seem to be connected with the extremely low carrier concentration of 10$^{-3}$ electron/Ce.  The Fermi energy of the electron carriers is estimated to be $\sim$20 K as described above, while that of the hole carriers could be one order-of-magnitude smaller because of the larger mass.  The suppression of $T_{\text{coh}}$ probably relates to these fine energy scales.  The huge $A$ does not conform to the Kadowaki-Woods relation, {\it A}/$\gamma$$^2$ $\sim$ 1$\times$10$^{-5}$ $\mu$$\Omega$cm(molK/mJ)$^2$, which is observed in many heavy fermions.\cite{Kadowaki86}  In the theoretical derivation of this relation,\cite{Takimoto96} metals with large carrier concentration ($\sim$1 electron/Ce) are assumed.  This is, however, not the case with CeNiSn, in which the formation of the pseudogap suppresses $\gamma$, while it enhances resistivity through reduced carrier number.  Thus the anomalously large {\it A} can be qualitatively explained.

One might argue that electron-phonon scattering could also lead to $T^2$ behavior in semimetals.  For example, the resistivity of bismuth varies as $T^2$ between 0.3 and 1 K.\cite{Uher77}  The essence of the electron-phonon scenario of the $T^2$ resistivity is that, if the FS is small and anisotropic, large-angle scattering (i.e., scattering angle $\sim$ 180$^{\circ}$) by phonons may survive down to very low temperatures:\cite{Kukkonen78}  At a low temperature {\it T}, most of excited phonons have wave numbers smaller than $q_T=k_BT/\hbar s$, where $s$ is a sound velocity.  If a minimal caliper of a FS is smaller than $q_T$, appreciable large-angle scattering can still occur and the temperature dependence of resistivity remains weaker than $T^5$ that is expected from the Bloch-Gr\"uneisen law.\cite{Ziman64}  Specifically, in the case of a cylindrical FS, it can be shown that the resistivity exhibits $T^2$ dependence in an extended temperature range before the $T^5$ dependence appears.\cite{Kukkonen78}  This scenario, however, does not apply to CeNiSn.  The sound velocity in CeNiSn is $2\sim4\times10^5\text{ cm/s}$,\cite{Nakamura91, Suzuki92} yielding $q_T=1\sim2\times10^{-4}\text{ \AA}^{-1}$ at $T=0.035{\text K}$, the lowest temperature of the measurements.  The observed SdH frequencies, $\sim$100 T, correspond to a FS cross-section of $\sim1\times10^{-2}\text{ \AA}^{-2}$.  For simplicity, let us assume the cross-section to be rectangular.  The shorter sides of the rectangle must be comparable to $q_T$ so that large-angle scattering can be effective.  Accordingly, the longer sides will be $100\sim50\text{ \AA}^{-1}$.  Such extremely anisotropic FS (aspect ratio $>$ 10$^5$) can never be compatible with the anisotropy in the resistivity: the ratio of the resistivities along the most resistive {\it c} and least resistive {\it a} axes is $\sim$4 at most.\cite{Nakamoto95}

Lastly, we discuss magnetic-field effects on the electronic structure.  The negative magnetoresistance and nonlinear Hall resistivity for {\it B} $\parallel$ {\it a} [Figs.~\ref{fig:1}(c) and (e)] suggest that the electronic structure changes with the field, as was previously pointed out.\cite{Takabatake92, Takabatake96}  Although the magnetoresistance maximum and bend (or dip) can not necessarily be identified with phase transitions, it seems reasonable to regard them as characteristic fields of the field-induced change.  The magnetoresistance maximum probably signals the onset of substantial changes.  This view is supported by the fact that the Hall resistivity for {\it B} $\parallel$ {\it c} deviates from nearly linear field dependence, i.e., normal behavior, at the field of the magnetoresistance maximum.  On the other hand, it is not clear what underlies the bend (or dip).  These characteristic fields increase as the field is tilted away from the {\it a} axis.  This may be related to that the magnetic susceptibilities along the {\it b} and {\it c} axes are less than half of that along the {\it a} axis.\cite{Takabatake90, Nakamoto95} 

The sudden appearance of the high-frequency oscillations above the magnetoresistance dip (Fig.~\ref{fig:5}) suggests that a larger FS appears at high fields as a result of the changes in the electronic structure.  The quasiparticle mass associated with the high frequency is no smaller than those found for the low-frequency branch.  These are basically consistent with that $\gamma$ is enhanced in fields along the {\it a} axis.\cite{Takabatake92, Izawa96}  The variation of $\gamma$ is gradual and featureless, compared to the sudden appearance of the high frequency oscillation.  The discrepancy could be ascribed to the difference in the field direction and/or the fact that $\gamma$ was measured on an early semiconducting sample (Ref.~\onlinecite{Takabatake92}) or at a much higher temperature of 3.3 K.\cite{Izawa96}

\section{Conclusion}

We have established that a small number of intrinsic charge carriers with enhanced mass do exist in the ground state of CeNiSn.  The carrier concentration is estimated to be $\sim10^{-3}$ electron / Ce, while the quasiparticle mass is $\sim$10-20 $m_e$.  Combining these two parameters, we have estimated the contribution of the observed FS to $\gamma$ to be a few mJ / molK$^2$, which is one order-of-magnitude smaller than the experimental $\gamma$.  Thus, the essential question raised by the two-fluid model still remains to be solved; whether or not neutral spin excitations, besides charge carriers, are necessary to explain low-temperature thermodynamic properties in CeNiSn.  The resistivity exhibits $T^2$-dependence below $T_{\text{coh}}=0.16\text{ K}$.  Viewing this dependence as a Fermi-liquid behavior, we have argued that the low $T_{\text{coh}}$ and huge $A$ associated with it may be related to the extremely low carrier concentration.

We have observed that the high-frequency SdH oscillations appear above the magnetoresistance-dip anomaly for certain field directions and have discussed these observations in terms of field-induced changes in the electronic structure.  We note that the Zeeman energy at 20 T is $\sim$20 K for a moment of $\sim$1 $\mu$$_B$ and is comparable to (or much larger than) the Fermi energy of the electron (hole) carriers derived in the present work.  This suggests that competition between these energies may lead to fundamental change of the electronic structure.  Therefore field-effects on the electronic structure deserve further studies.

\begin{acknowledgments}
Work at Hiroshima University was supported by a Grant-in-Aid for Scientific Research (COE Research 13CE2002) of MEXT, Japan.
\end{acknowledgments}

\end{document}